\documentclass[runningheads]{llncs}
\usepackage[T1]{fontenc}
\usepackage{graphicx,verbatim, amsmath, url}
\begin{document}
\title{Explainability in mulimodal deep transformation models for stroke outcome prediction}
\titlerunning{Explainability in multimodal DTMs}
% If the paper title is too long for the running head, you can set
% an abbreviated paper title here
%

\author{Lisa Herzog\inst{1}%\orcidID{0009-0000-2095-9516} 
\and
Jonas Brändli\inst{2} \and
Maurice Schneeberger\inst{3} \and
Loran Avci\inst{2} \and
Nordin Dari\inst{2} \and
Martin Hänsel\inst{1} \and
Hakim Baazaoui\inst{1} \and
Pascal Bühler\inst{2,4} \and
Susanne Wegener\inst{1} \and
Beate Sick\inst{2,3,4}
}
%index{Herzog, Lisa}
%index{Brändli, Jonas}
%index{Schneeberger, Maurice}
%index{Avci, Loran}
%index{Dari, Nordin}
%index{Hänsel, Martin}
%index{Baazoui, Hakim}
%index{Bühler, Pascal}
%index{Wegener, Susanne}
%index{Sick, Beate}

%
\authorrunning{L. Herzog et al.}
% First names are abbreviated in the running head.
% If there are more than two authors, 'et al.' is used.
%
% @Beate: Email von corresponding author (du oder ich)?
\institute{University Hospital Zurich, Department of Neurology, Frauenklinikstrasse 26, 8091 Zurich, Switzerland\\ \and
Zurich University of Applied Sciences, Institute of Data Analysis and Process Design (IDP), Technikumstrasse 9, 8401 Winterthur, Switzerland\\ \and
University of Zurich, Epidemiology, Biostatistics and Prevention Institute (EBPI), Hirschengraben 84, 8001 Zurich, Switzerland \and
Thurgau Institute for Digital Transformation (TIDIT), Hauptstrasse 21a, 8280 Kreuzlingen, Switzerland \email{beate.sick@uzh.ch, beate.sick@tidit.ch}}

\maketitle              % typeset the header of the contribution
\begin{abstract}
Multimodal prediction models based on imaging and clinical data are increasingly used for clinical decision support, yet their interpretability remains limited. We present multimodal Deep Transformation Models (DTMs) combining statistical approaches and neural networks to achieve strong predictive performance while preserving interpretability for tabular data. A key contribution of this work is the adaption of the xAI methods Grad-CAM and Occlusion to DTMs relying on 3D CNNs, enabling interpretation of the image branch through the generation of explanation maps. We developed DTMs to predict functional independence three months after stroke using diffusion-weighted imaging and clinical data from 407 patients. In a ten-fold cross-validation, the models achieved state-of-the-art predictive performance (AUC 0.81 [0.75, 0.87]) while maintaining interpretability for tabular features, with functional independence before stroke and stroke severity on admission emerging as the strongest predictors. Explanation maps from both xAI methods highlighted consistent regions, including frontal lobe areas which are known to be associated with age, a strong predictor of functional outcome. Notably, these regions disappeared once age was included as an explicit tabular predictor. Similarity analyses of explanation maps revealed distinct spatial patterns, providing meaningful insights into stroke pathophysiology, systematic error analysis and hypothesis generation.
\end{abstract}
\section{Introduction}

Accurate and trustworthy outcome prediction models are increasingly important for clinical decision support, yet building such models remains challenging due to the multimodal nature of patient data and strict requirements on model transparency. Clinically suitable models should jointly integrate heterogeneous sources such as neuroimaging and structured clinical data, achieve strong predictive performance, and provide interpretable results that allow clinicians to understand how predictions arise \cite{Rudin2019}. Classical statistical models offer interpretability through interpretable parameter estimates but struggle with high-dimensional inputs such as medical images, whereas modern deep learning (DL) achieves strong performance at the cost of limited explainability. 

Deep Transformation Models (DTMs) combine statistical modeling with neural networks (NNs) to enable interpretable multimodal prediction \cite{KookHerzog2020,baumann2021deep}. They estimate a conditional probability distribution of the outcome while integrating different input modalities with (potentially deep) NNs. Their design preserves interpretability through interpretable parameter estimates for tabular features while allowing complex feature extraction from imaging data. Prior work has demonstrated strong predictive performance of DTMs in medical applications \cite{herzog2023deep,Campanella2022_DeepTrafoSurvival}. However, the image branch remains difficult to interpret.

Post-hoc explainable artificial intelligence (xAI) methods provide a practical solution to the black box problem. Without altering the architecture, they highlight the contribution of image regions to a model's output. While such explanations do not render models inherently transparent and may have limitations regarding robustness \cite{Rudin2019,VanderVelden2022XAI}, they currently represent the most practical approach for explainability. However, their application to probabilistic DTMs has not been established yet.

In this work, we extended Occlusion \cite{zeiler2014visualizing} and Grad-CAM \cite{selvaraju2017grad} to multimodal DTMs with 3D CNN image branches. Using a cohort of 407 stroke and transient ischemic attack (TIA) patients, we evaluated predictive performance, interpretability of clinical parameters and systematically analyzed explanation maps to identify characteristic spatial patterns associated with functional outcome prediction. This establishes a general framework for explainability in multimodal transformation models and supports hypothesis generation and error analysis in clinical prediction tasks.

\section{Material and methods}

\subsection{Data} 
We analyzed a retrospectively collected cohort of 407 patients with ischemic stroke (n=295) or TIA (n=112) admitted to the University Hospital Zurich between 2014 and 2018. TIA and stroke patients frequently present with similar neurological symptoms, making neuroimaging essential for diagnosis and characterization of tissue injury. All patients underwent diffusion-weighted imaging (DWI) within three days after symptom onset, with most scans acquired within the first 24 hours. Brain volumes were preprocessed to a size of 128x128x28 and normalized to have zero mean and unit standard deviation. Tabular data included demographics, vascular risk factors and admission information (see Fig.~\ref{fig:beta_values}). Categorical variables were dummy-encoded and numerical variables standardized to ensure comparability of parameter estimates. Functional outcome was assessed using the modified Rankin Scale (mRS) at three months, dichotomized into favorable (mRS 0-2, n=332) and unfavorable (mRS 3-6, n=75) outcome. Ethical approval was obtained from the local ethics committee and all participants have provided written informed consent.

\subsection{Deep Transformation Models}

DTMs are interpretable probabilistic models for multimodal prediction estimating a conditional probability distribution $F_{Y}(y|\textsf{B},x)$ of an outcome $Y$ given imaging data $\textsf{B}$ and tabular features $x$ \cite{KookHerzog2020}. Instead of estimating the outcome distribution directly, DTMs estimate a transformation function $h$ that maps a predefined latent distribution $F_Z$ to the targeted outcome distribution, such that
\begin{equation}
    F_Y(y|\textsf{B},x) = F_Z(h(y|\textsf{B},x)).
\end{equation}
The parameter-free probability distribution $F_Z$ defines the interpretational scale of the model parameters. For instance, using the standard logistic distribution $\sigma(z)=\frac{1}{1+\exp(-z)}$ yields a formulation analogous to logistic regression, allowing additive model components to be interpreted on the log-odds scale \cite{Hothorn2014conditional,KookHerzog2020}.

For binary outcomes, the transformation function represents a threshold on the z-scale, partitioning $F_Z(z)$ into two segments which determine the class probabilities $p_{0} = P(y_0|\textsf{B},x) = \sigma(h(y_0|\textsf{B},x))$ and $p_{1} = P(y_1|\textsf{B},x) = 1 - p_0$ (see Fig.~\ref{fig:ex_framework}). Here, $y_0$ represents a favorable, $y_1$ an unfavorable outcome. We define 
\begin{equation}
    h(y_0|\textsf{B},x)=\vartheta_0(\textsf{B})-x^\top\beta,
\end{equation}
where $\vartheta_0(\textsf{B})$ is estimated by a 3D CNN processing imaging data and $x^\top\beta$ with a NN that represents a linear shift for tabular features. The NNs are jointly fitted by minimizing the negative log-likelihood 
\begin{equation}
    \text{NLL}=-\frac{1}{n}\sum_{i=1}^{n}\log(F_Z(y_{k}|\textsf{B},x)-F_Z(y_{k-1}|\textsf{B},x))
\end{equation}
using standard stochastic optimization. With $F_Z(z)=\sigma(z)$, $\exp(\beta_k)$ describes the multiplicative change in the odds of a favorable outcome associated with a one unit increase in $x_k$, when holding the remaining inputs constant. For a detailed discussion we refer to \cite{KookHerzog2020}.

\subsection{Explanation maps}

\begin{figure}[htbp]
\centering
\includegraphics[width=0.9\textwidth]{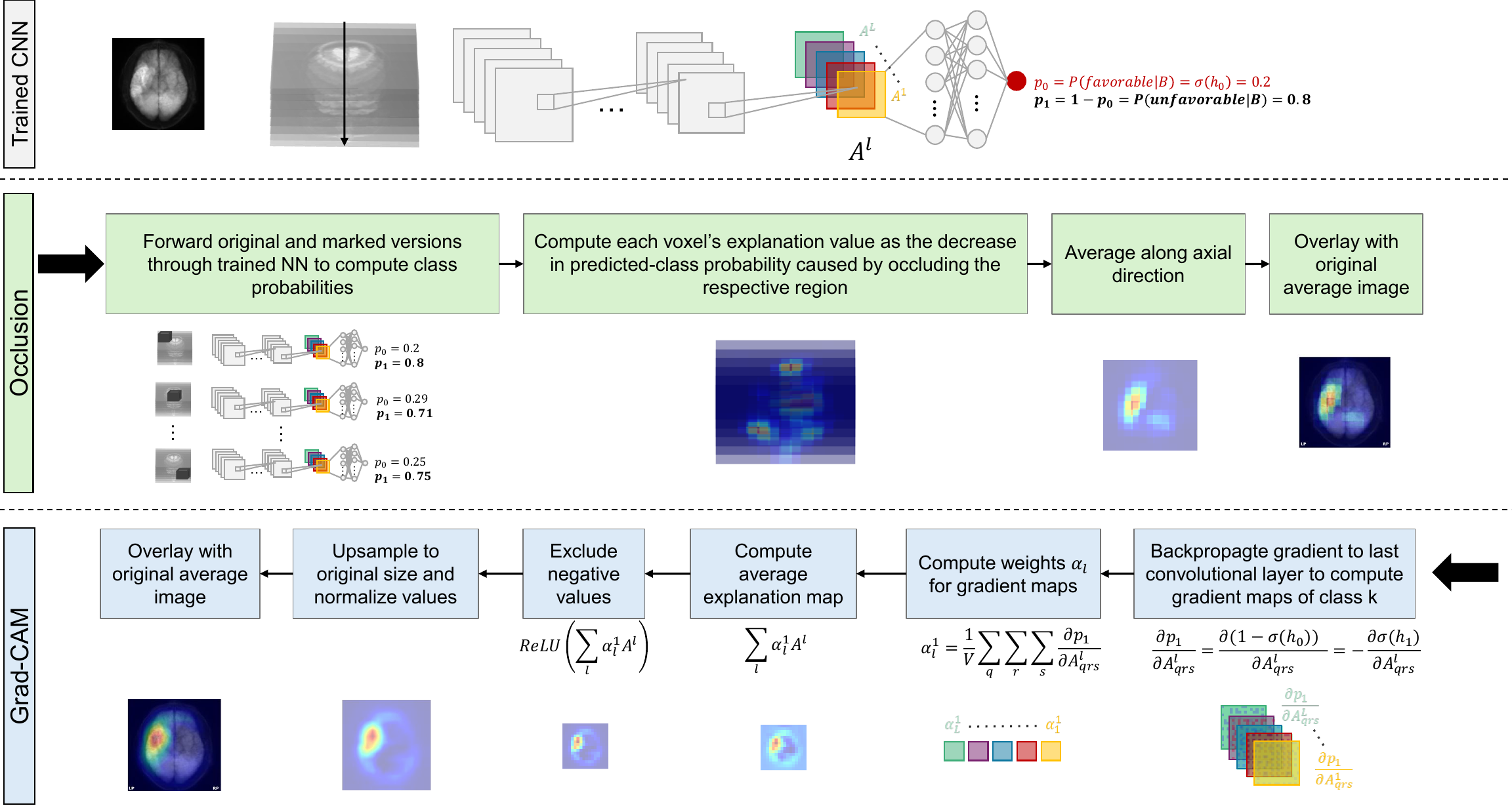}
\caption{Overview of the Grad-CAM and Occlusion method for DTMs. Both approaches build on the trained CNN. Occlusion identifies relevant regions by systematically masking parts of the input and measuring the resulting change in the probability of the predicted class. Grad-CAM derives voxel importance from the gradients flowing from the predicted class back to the last convolutional layer.}
\label{fig:occlusion_gradcam}
\end{figure}

To obtain visual explanations of the image branch, we adapted Occlusion \cite{zeiler2014visualizing} and Grad-CAM \cite{selvaraju2017grad} to the DTM framework (see Fig.~\ref{fig:occlusion_gradcam}). Both methods operate on the trained CNN and account for the transformation function output. Grad-CAM provides insight into spatial features represented in the network, whereas Occlusion directly evaluates the effect of localized perturbations on the probability for the predicted class, therefore, providing complementary perspectives.

\subsubsection{Occlusion} 
Occlusion was performed by systematically masking regions of the 3D input and measuring changes in the probability of the predicted class. Regions whose masking decreased the predicted probability were considered important for the decision. For the 3D images, we used an occlusion size of 18x18x4 and a stride width of 10x10x3.

\subsubsection{Grad-CAM}
In standard 2D CNN classifiers, Grad-CAM leverages gradient information flowing from the output node for the predicted class $k$ into the last convolutional layer. In DTMs, however, the 3D CNN contributes to the transformation function rather than directly producing class probabilities. Considering $h_0 = h(y_0|B,x) = \vartheta_0(B)-x^T\beta$, we therefore quantify the importance of each 3D activation map $A^{l}$, $l = 1,\dots,L$ in the last convolutional layer by computing the gradient for each voxel $A^{l}_{(q,r,s)}$
\begin{equation}
\frac{\partial p_1}{\partial A^{l}_{(q,r,s)}}
%=\frac{\partial (1- \sigma(h_0))}{\partial A^{l}_{(q,r,s)}}
=\frac{\partial \sigma(\vartheta_{0}(\textsf{B}) - x^{T}\beta)}{\partial A^{l}_{(q,r,s)}}
%=\frac{\partial \sigma(\vartheta_{0}(\textsf{B}))}{\partial A^{l}_{(q,r,s)}}
= \sigma(\vartheta_{0}(\textsf{B}))(1-\sigma(\vartheta_{0}(\textsf{B})))  \frac{\partial\vartheta_{0}(\textsf{B})}{\partial A^{l}_{(q,r,s)}}.  
\end{equation}
The last expression only involves differentiable components while the gradient of $\vartheta_{0}(\textsf{B})$ is calculated automatically via backpropagation. Pooling these gradients yields map weights
\begin{equation}
\alpha_l^1 = \frac{1}{R\cdot S}  \sum_{q}  \sum_{r} \sum_{s} \frac{\partial p_1}{\partial A_{qrs}^l}.
\end{equation}
The final explanation map is obtained via a weighted combination of activation maps followed by ReLU and upsampling
\begin{equation}
L_1^{\text{Grad-CAM}} = \text{ReLU}\left(\sum_{l} \alpha_l^1 \cdot A^l\right).
\end{equation}
This formulation enables gradient-based explanation maps consistent with the probabilistic DTM structure. Derivations for other transformation functions are equivalent.

\subsection{Experiments}

\begin{figure}
\centering
\includegraphics[width=0.9\textwidth]{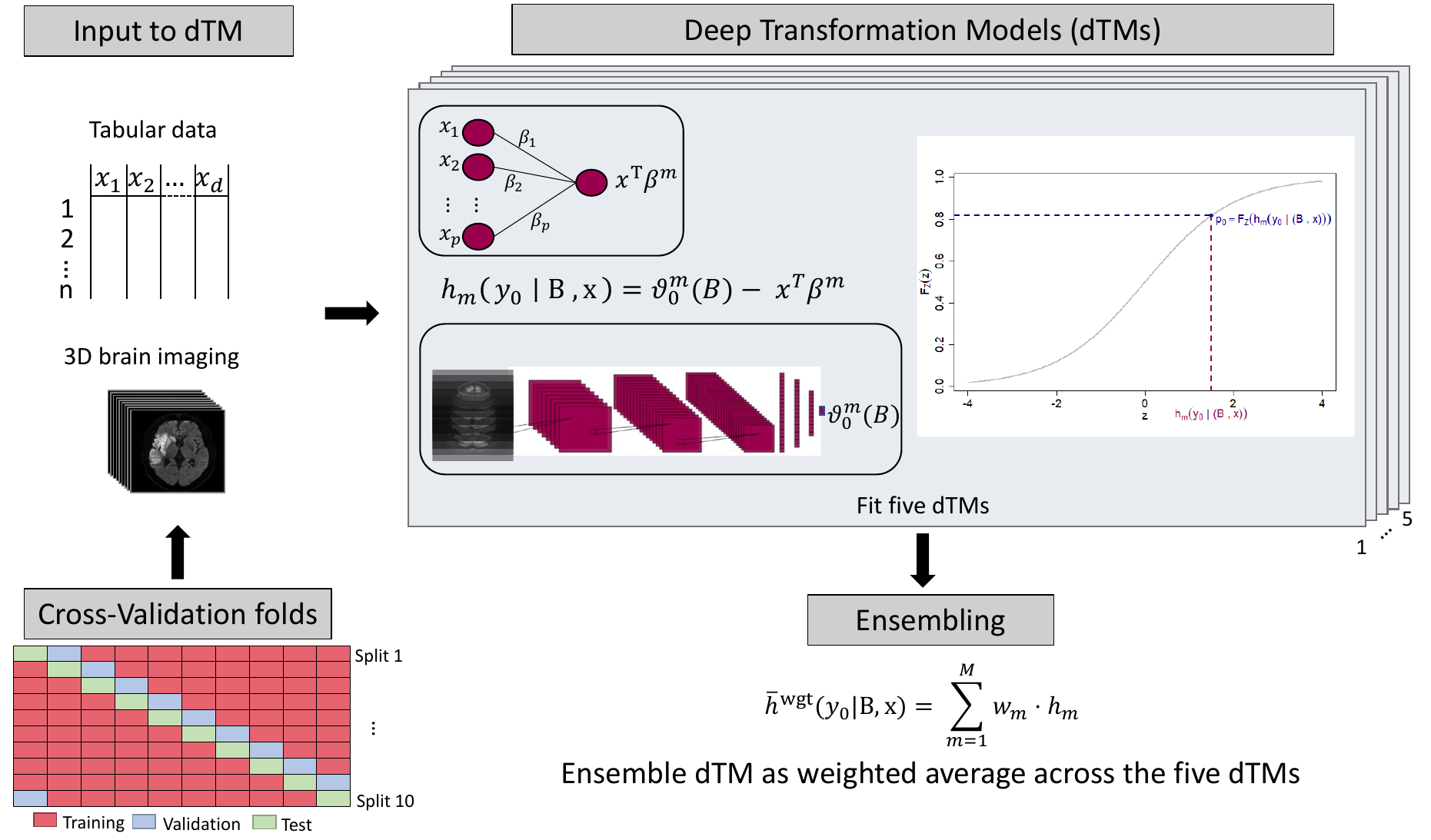}
\caption{DTM training and evaluation setup. In each CV fold, we fit $M=5$ DTMs on the training data and determine a threshold on the validation set to separate the standard logistic distribution $F_Z(z)=\sigma(z)$ into two segments to yield a probability for favorable $p_{0} = \sigma(h_m(y_0|\textsf{B},x))$ vs.~unfavorable ($1-p_0$) outcome. The final ensemble DTM threshold is a weighted average of the $M$ thresholds.}
\label{fig:ex_framework}
\end{figure}

We compared an imaging-only model with a complex intercept (\textbf{CI\textsubscript{B}}; $h(y_{0}|\textsf{B}) = \vartheta_{0}(\textsf{B})$), a tabular-only model with simple intercept and linear shift (\textbf{SI-LS\textsubscript{x}}; $h(y_0|x) = \vartheta_{0} + x^{T}\beta$; logistic regression), and a multimodal model (\textbf{CI\textsubscript{B}-LS\textsubscript{x}}; $h(y_0|\textsf{B},x) = \vartheta_{0}(\textsf{B}) + x^{T}\beta$). In addition, we fitted a null model to assess if the models have learned meaningful features to enable interpretability of explanation maps. The linear shift terms were implemented with a fully connected, single-layer NN. The image branch of the models used a 3D CNN with four convolutional blocks (3x3x3 kernels, 32-32-64-64 filters), global average pooling and two fully connected layers (128 units, dropout 0.3). During training, data augmentation (zooming, flipping, shifting, rotating) was applied.

All models were evaluated using stratified ten-fold cross-validation (see Fig.~\ref{fig:ex_framework}). For the models including imaging data, deep ensembling ($M=5$) \cite{kook2026trafoensemble} was applied, resulting in five transformation functions per fold that were averaged with weights $w_m$ optimized to minimize the NLL on the validation data with $\bar{h}^{wgt} = \sum_{m=1}^{M} w_m h_m$. 
The five 3D explanation maps resulting from Occlusion and Grad-CAM were averaged using the same weights $w_m$. To produce a final, interpretable 2D explanation map, the 3D ensemble map was overlayed with the original 3D brain volume and averaged along the axial direction (see Fig.~\ref{fig:occlusion_gradcam}).

Test performance was assessed using NLL, AUC, accuracy, specificity and sensitivity. For specificity, sensitivity, and accuracy, we report 95\% Wilson confidence intervals (CIs). CIs for NLL and AUC were obtained via bootstrapping. Class labels were derived from predicted probabilities using thresholds determined to maximize the geometric mean of the true-positive and true-negative rates on the validation set of each fold. All code is publicly available at \url{https://github.com/liherz/xAI_paper}.

Similarity plots of Grad-CAM explanation maps were used to identify spatial patterns, understand stroke pathophysiology and perform error analysis and hypothesis generation. Therefore, we applied t-SNE on the features of the explanation maps extracted from the last convolutional layer of a ResNet-50 trained on Imagnet \cite{He2015}. An interactive plot is available under \url{https://liherz.shinyapps.io/tsne_xAI_shinyio/}. 

% ----------------------------------------------------------------------------------------
\section{Results}

\begin{table}[htbp]
\centering
\caption{Predictive performance of models using diffusion weighted imaging (DWI) data, tabular clinical data (Clinical) and a combination of both. NLL = Negativ log-likelihood, AUC = Area under the Receiver Operating Characteristic Curve.}%
  {\begin{tabular}{lllll}
  \hline
 & \bfseries DWI & \bfseries Clinical & \bfseries DWI + Clinical \\ 
 \hline
\bfseries NLL 
% & 0.478 [0.423,  0.536] SI
& 0.452 [0.379, 0.528] 
& 0.368 [0.308, 0.433]
& 0.384 [0.306, 0.470] \\ % \bf{0.381 [0.313, 0.456]} 
\bfseries  AUC     
% & 0.498 [0.430, 0.570]  SI
& 0.714 [0.649, 0.777] 
& 0.809 [0.747, 0.867]
& 0.812 [0.752, 0.868] \\ % \bf{0.793 [0.73\phantom{0}, 0.853]} \\ %
\bfseries Specificity 
% & 1.0\phantom{00} [0.989, 1.0\phantom{00}] SI
& 0.792 [0.745, 0.832]
& 0.783 [0.736, 0.824]
& 0.795 [0.749, 0.835] \\ %0.801 [0.755, 0.841] \\ 
\bfseries Sensitivity 
% &  0.0\phantom{00} [0.0\phantom{00}, 0.049] SI
& 0.480 [0.371, 0.591]
& 0.627 [0.514, 0.727]
& 0.640 [0.527, 0.739] \\ % 0.627 [0.514, 0.727] \\ 
\bfseries Accuracy    
% & 0.816 [0.775, 0.850] SI
& 0.735 [0.690, 0.775] 
& 0.754 [0.710, 0.794]
& 0.767 [0.723, 0.805] \\ % 0.769 [0.726, 0.807] \\
\hline
\end{tabular}}
\label{tab:outcome_metrics_update}
\end{table}

The multimodal model achieved the highest predictive performance (AUC 0.812 [0.752, 0.868], see Table~\ref{tab:outcome_metrics_update}). However, the model using clinical data alone performed similar (AUC 0.809 [0.747, 0.867]), indicating that clinical features contain most of the predictive information. All models showed higher specificity (correctly predicted favorable outcomes), likely reflecting the predominance of favorable outcomes in the cohort. Comparison with the null model (AUC 0.498 [0.430, 0.570]) confirms that all models have learned meaningful patterns, supporting a cautious but informative interpretation of the explanation maps.

\begin{figure}[t]
\centering
\includegraphics[width=0.6\textwidth]{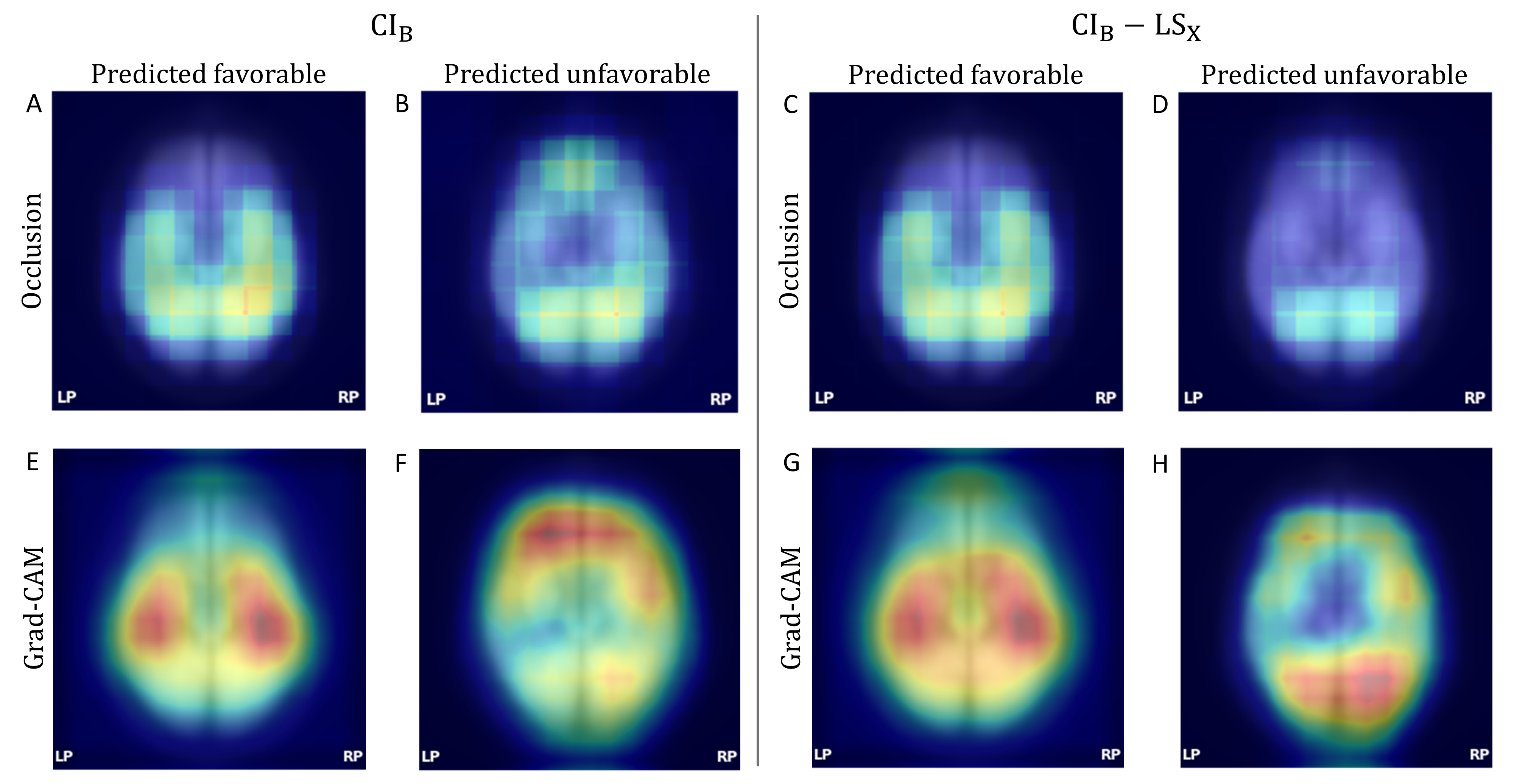}
\caption{Average explanation maps. The figure shows the average across the explanation maps for patients predicted having a favorable or unfavorable outcome, produced with the CI\textsubscript{B} and CI\textsubscript{B}-LS\textsubscript{X} model when applying Occlusion and Grad-CAM.}
\label{fig:explanationmap_average_CIBLSX}
\end{figure}

Occlusion and Grad-CAM highlighted similar regions (see Fig.~\ref{fig:explanationmap_average_CIBLSX}). However, Grad-CAM yielded smoother and more distinct average explanation maps while being less computationally expensive. For unfavorable outcome prediction, the imaging-only model \textbf{CI\textsubscript{B}} focused on frontal lobe regions, which are known to be associated with age, a strong predictor of functional outcome \cite{fujita2023characterization}. These regions disappeared in the multimodal \textbf{CI\textsubscript{B}-LS\textsubscript{x}} model, indicating that the imaging-only model had captured age-related patterns to predict unfavorable outcomes that became redundant when age was included explicitly as a tabular predictor.

\begin{figure}[t]
\centering
\begin{minipage}{0.48\textwidth}
    \centering
    \includegraphics[width=\textwidth]{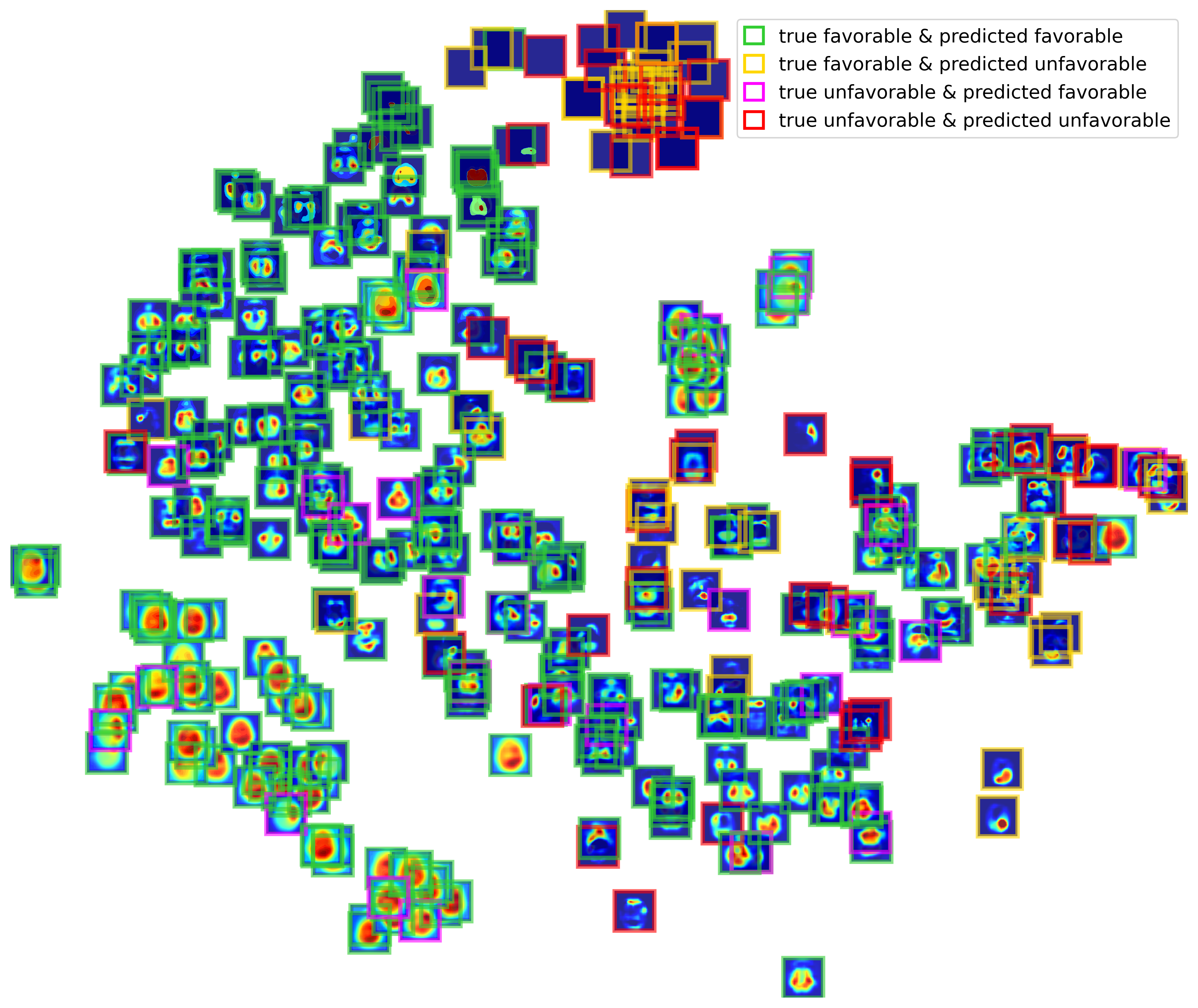}
\end{minipage}
\hfill
\begin{minipage}{0.48\textwidth}
    \centering
    \includegraphics[width=\textwidth]{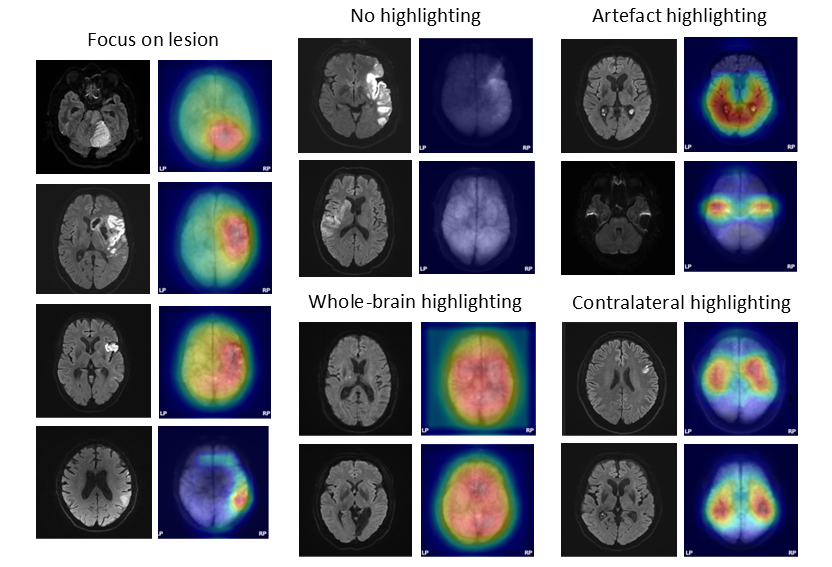}
\end{minipage}
\caption{Grad-CAM explanation analysis. (A) Similarity structure of explanation maps visualized using t-SNE, illustrating clustering behavior across correctly and incorrectly predicted favorable and unfavorable outcome groups. (B) Distinct spatial patterns of averaged explanation maps with representative volumetric examples.}
\label{fig:tsne}
\end{figure}

Cluster analysis revealed several consistent patterns (see Fig.~\ref{fig:tsne}). In most stroke cases, the model focused on lesion areas appearing as hyperintensities on DWI. Occasionally, imaging artifacts were highlighted (e.g. ventricular hyperintensities). Frequently, the model highlighted bilateral regions, suggesting assessment of hemispheric/contralateral symmetry. When strong clinical predictors (e.g. high NIHSS scores) were available, Grad-CAM maps often showed minimal activation, indicating reduced reliability on images. In patients with TIA or very small lesions, when only a few slices show visible abnormalities, activation was on the entire image, suggesting that the image branch contributed global rather than localized information.

\begin{figure}[t]
\centering
\begin{minipage}{0.48\textwidth}
    \centering
    \includegraphics[width=\textwidth]{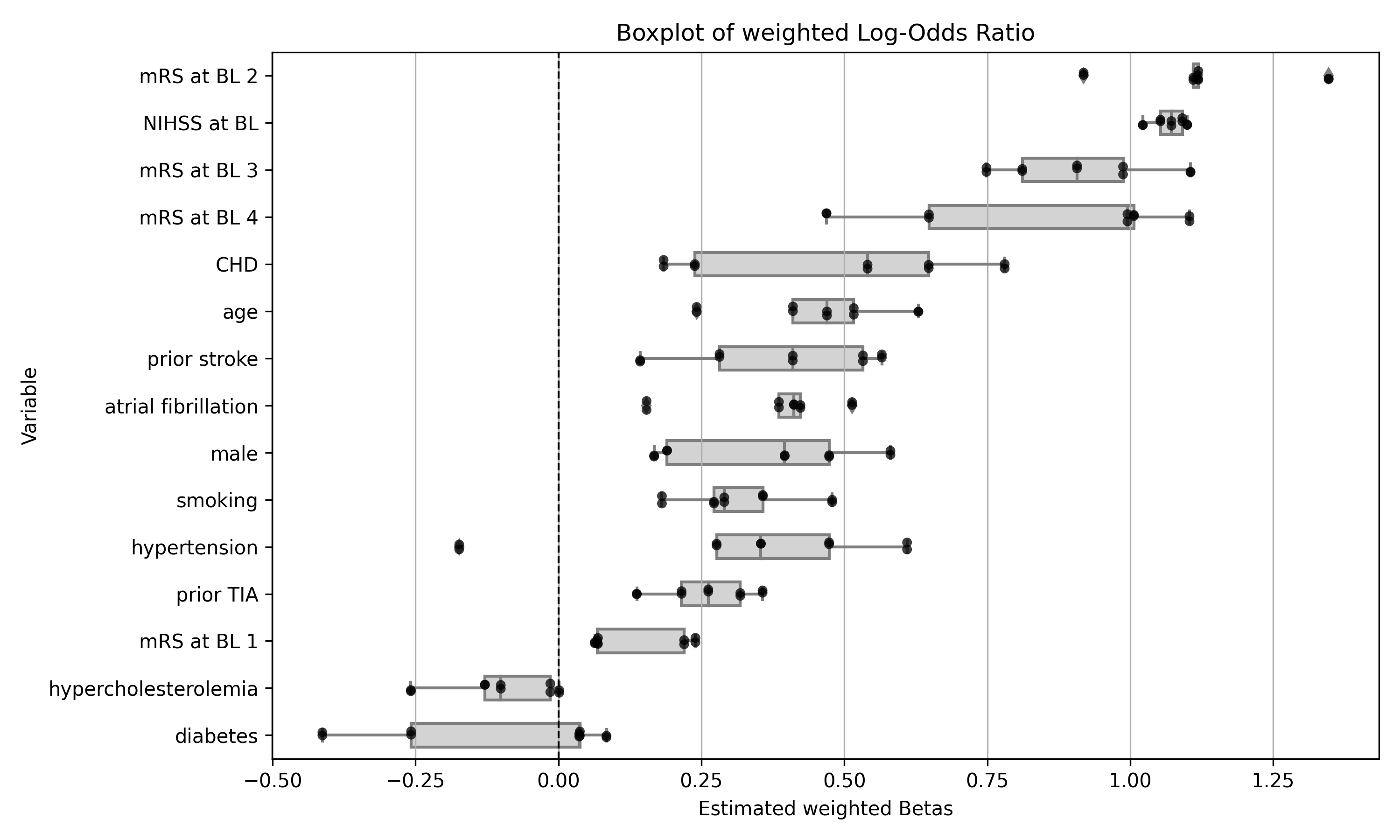}
\end{minipage}
\hfill
\begin{minipage}{0.48\textwidth}
    \centering
    \includegraphics[width=\textwidth]{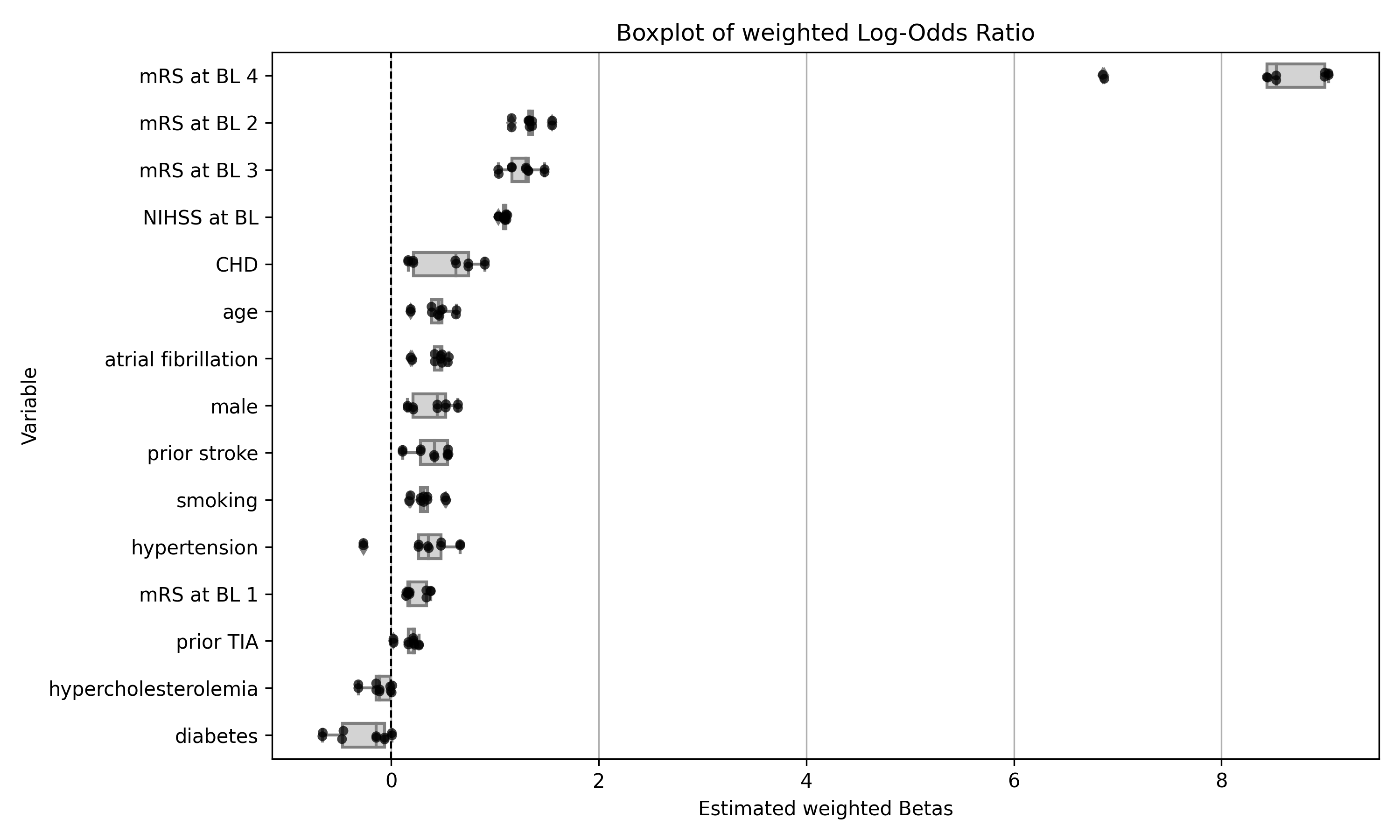}
\end{minipage}
\caption{Estimated log-odds ratios with 95\% bootstrap confidence intervals for tabular predictors derived from (A) the SI-LS\textsubscript{X} model and (B) the tabular component of the CI\textsubscript{B}-LS\textsubscript{X} model. Estimates were obtained from 10-fold cross-validation and averaged across ensemble members. Categorical features (sex, smoking, hypertension, prior stroke or TIA, atrial fibrillation, coronary heart disease (CHD), diabetes, hypercholesterolemia, pre-stroke mRS (mRS at BL)) are shown with respect to the reference category. Parameter estimates of numeric variables (age, National Institutes of Health Stroke Scale on admission (NIHSS at BL)) are based on the standardized measures.}
\label{fig:beta_values}
\end{figure}

Beyond imaging explanations, DTMs provide interpretable parameters for tabular predictors when being included as linear shift terms (see Fig.~\ref{fig:beta_values}). Here, pre-stroke mRS and admission NIHSS emerged as the strongest predictors in models with and without imaging data. 

% ----------------------------------------------------------------------------------------
\section{Discussion}

In this work, we extended Occlusion and Grad-CAM to DTMs with multimodal input, enabling post-hoc interpretability for models integrating 3D neuroimaging and clinical data. Using these models, we predicted favorable vs.~unfavorable functional outcomes in TIA and stroke patients with state-of-the-art performance. We identified specific imaging patterns relevant for prediction and provided interpretable parameters for clinical features, with NIHSS on admission and pre-stroke mRS emerging as the most influential predictors. Although sample size was limited, cross-validation ensured robustness, and results were consistent with clinical knowledge and literature. Prediction performance was comparable to other multimodal prediction models for acute ischemic stroke \cite{Klug2024,Liu2023,Oliveira2023}. Explanation maps from Grad-CAM and Occlusion highlighted similar regions, with Grad-CAM producing smoother, more visually coherent maps while requiring less computational power making it the preferred approach. Both xAI methods offered meaningful insights into stroke pathophysiology and highlighted characteristic patterns that may indicate false predictions, supporting systematic error analysis.

In conclusion, DTMs combine the strengths of deep learning and statistical modeling, providing reliable and interpretable predictions for multimodal data. By adapting explanation maps to this framework, the image branch becomes more transparent, supporting hypothesis generation and assessment of model trustworthiness, which is critical for clinical translation.

\paragraph{The paper is accpted at MICCAI 2026}

\bibliographystyle{splncs04}
\bibliography{bibliography}

@article{fujita2023characterization,
  title={Characterization of brain volume changes in aging individuals with normal cognition using serial magnetic resonance imaging},
  author={Fujita, Shohei and Mori, Susumu and Onda, Kengo and Hanaoka, Shouhei and Nomura, Yukihiro and Nakao, Takahiro and Yoshikawa, Takeharu and Takao, Hidemasa and Hayashi, Naoto and Abe, Osamu},
  journal={JAMA Network Open},
  volume={6},
  number={6},
  pages={e2318153--e2318153},
  year={2023},
  publisher={American Medical Association}
}

@inproceedings{baumann2021deep,
  title={Deep conditional transformation models},
  author={Baumann, Philipp FM and Hothorn, Torsten and R{\"u}gamer, David},
  booktitle={Joint European Conference on Machine Learning and Knowledge Discovery in Databases},
  pages={3--18},
  year={2021},
  organization={Springer}
}

@Article{Campanella2022_DeepTrafoSurvival,
  author  = {Campanella, Gabriele and Kook, Lucas and Häggström, Ida and Hothorn, Torsten and Fuchs, Thomas J.},
  journal = {arXiv:2210.11366},
  title   = {Deep conditional transformation models for survival analysis},
  year    = {2022},
}

@Article{He2015,
  author  = {He, Kaiming and Zhang, Xiangyu and Ren, Shaoqing and Sun, Jian},
  journal = {IEEE International Conference on Computer Vision},
  title   = {Delving deep into rectifiers: Surpassing human-level performance on ImageNet classification},
  year    = {2015},
  doi     = {10.1109/ICCV.2015.123},
}

@article{herzog2023deep,
  title={Deep transformation models for functional outcome prediction after acute ischemic stroke},
  author={Herzog, Lisa and Kook, Lucas and G{\"o}tschi, Andrea and Petermann, Katrin and H{\"a}nsel, Martin and Hamann, Janne and D{\"u}rr, Oliver and Wegener, Susanne and Sick, Beate},
  journal={Biometrical Journal},
  volume={65},
  number={6},
  pages={2100379},
  year={2023},
  publisher={Wiley Online Library}
}

@article{Hothorn2014conditional,
author = {Hothorn, Torsten and Kneib, Thomas and B{\"{u}}hlmann, Peter},
doi = {10.1111/rssb.12017},
eprint = {1201.5786},
journal = {Journal of the Royal Statistical Society. Series B: Statistical Methodology},
number = {1},
pages = {3--27},
publisher = {[Royal Statisticapdl Society, Wiley]},
title = {{Conditional transformation models}},
volume = {76},
year = {2014}
}

@Article{KookHerzog2020,
  author  = {Kook \& Herzog, Lucas Lisa and Hothorn, Torsten and D{\"u}rr, Oliver and Sick, Beate},
  journal = {Pattern Recognition},
  title   = {Deep and interpretable regression models for ordinal outcomes},
  year    = {2022},
  issn    = {0031-3203},
  pages   = {108263},
  volume  = {122},
}

@Article{Klug2024,
  author  = {Julian Klug and Guillaume Leclerc and Elisabeth Dirren and Emmanuel Carrera},
  journal = {Communications Medicine},
  title   = {Machine learning for early dynamic prediction of functional outcome after stroke},
  year    = {2024},
  note    = {Published online: 13 November 2024},
  number  = {1},
  pages   = {232},
  volume  = {4},
  doi     = {10.1038/s43856-024-00666-w},
  url     = {https://doi.org/10.1038/s43856-024-00666-w},
}

@Article{Liu2023,
  author  = {Yongkai Liu and Yannan Yu and Jiahong Ouyang and Bin Jiang and Guang Yang and Sophie Ostmeier and Max Wintermark and Patrik Michel and David S. Liebeskind and Maarten G. Lansberg and Gregory W. Albers and Greg Zaharchuk},
  journal = {Stroke},
  title   = {Functional Outcome Prediction in Acute Ischemic Stroke Using a Fused Imaging and Clinical Deep Learning Model},
  year    = {2023},
  note    = {Originally published 24 July 2023},
  number  = {9},
  volume  = {54},
  doi     = {10.1161/STROKEAHA.123.044072},
  url     = {https://doi.org/10.1161/STROKEAHA.123.044072},
}

@Article{Oliveira2023,
  author    = {Oliveira, G. and Fonseca, A.C. and Ferro, J. and Oliveira, A.L.},
  journal   = {Diagnostics},
  title     = {Deep Learning-Based Extraction of Biomarkers for the Prediction of the Functional Outcome of Ischemic Stroke Patients},
  year      = {2023},
  number    = {24},
  pages     = {3604},
  volume    = {13},
  doi       = {10.3390/diagnostics13243604},
  publisher = {MDPI},
  url       = {https://doi.org/10.3390/diagnostics13243604},
}

@Article{Rudin2019,
  author  = {Rudin, Cynthia},
  title   = {Stop explaining black box machine learning models for high stakes decisions and use interpretable models instead},
  journal = {Nature Machine Intelligence},
  year    = {2019},
  volume  = {1},
  pages = {206–215},
}

@inproceedings{selvaraju2017grad,
  title={Grad-CAM: Visual explanations from deep networks via gradient-based localization},
  author={Selvaraju, Ramprasaath R and Cogswell, Michael and Das, Abhishek and Vedantam, Ramakrishna and Parikh, Devi and Batra, Dhruv},
  booktitle={Proceedings of the IEEE international conference on computer vision},
  pages={618--626},
  year={2017}
}

@article{kook2026trafoensemble,
  author  = {Kook, Lucas and G{\"o}tschi, Andrea and Baumann, Philipp and Hothorn, Torsten and Sick, Beate},
  title   = {Deep interpretable ensembles},
  journal = {Neurocomputing},
  year    = {2026},
  volume  = {682},
  month   = jun,
  doi     = {10.1016/j.neucom.2026.133394},
  url     = {https://doi.org/10.1016/j.neucom.2026.133394}
}

@inproceedings{zeiler2014visualizing,
  title={Visualizing and understanding convolutional networks},
  author={Zeiler, Matthew D and Fergus, Rob},
  booktitle={Computer Vision--ECCV 2014: 13th European Conference, Zurich, Switzerland, September 6-12, 2014, Proceedings, Part I 13},
  pages={818--833},
  year={2014},
  organization={Springer}
}

@Article{VanderVelden2022XAI,
  author  = {van der Velden, Bas H. M. and Kuijf, Hugo J. and Gilhuijs, Kenneth G. A. and Viergever, Max A.},
  journal = {Medical Image Analysis},
  title   = {Explainable artificial intelligence (XAI) in deep learning-based medical image analysis},
  year    = {2022},
  month   = jul,
  volume  = {79},
  doi     = {10.1016/j.media.2022.102470},
}

\end{document}